\documentclass[prd,aps,floatfix,nofootinbib,preprint,tightenlines]{revtex4}
\usepackage{dcolumn}
\usepackage{amsmath}
\usepackage{latexsym}
\usepackage{graphicx}
\usepackage{bm}

\def\co{{\cal O}}

\def\bx{{\bf x}}

\def\svev#1{\left\langle #1\right\rangle}       

\def\Tr{{\rm Tr}\,}

\def\Re{{\rm Re\,}}

\long \def \blockcomment #1\endcomment{}
\def\det{{\rm det}}

\def\ket#1{\mathinner{|{#1}\rangle}}

\newcommand{\bee}{\begin{equation}}
\newcommand{\ee}{\end{equation}}
\newcommand{\beea}{\begin{eqnarray}}
\newcommand{\eea}{\end{eqnarray}}

\begin{document}
\title{
Lattice baryons in the $1/N$ expansion}
\author{Thomas DeGrand}%
 \affiliation{Department of Physics,
University of Colorado, Boulder, CO 80309, USA}

\begin{abstract}
Results are presented for hadron spectroscopy with gauge groups
$SU(N)$ with $N=3$, 5, 7. Calculations use the quenched approximation.
Lattice spacings are matched using the static potential. Meson spectra show independence on $N$ and
vacuum-to-hadron matrix elements scale as $\sqrt{N}$.
The baryon spectrum shows the excitation levels  of a rigid rotor.
\end{abstract}

\maketitle

\section{Introduction}
Replacing the ``3'' of color $SU(3)$ by ``N'''' and then taking N to infinity has a long
history in the (continuum) phenomenology of the strong interactions, 
dating back to Refs.~\cite{'tHooft:1973jz,'tHooft:1974hx}.

There is also a literature of lattice simulations applied to gauge theories
with the group $SU(N)$, for moderately
large $N$. Most of it
\cite{Lucini:2005vg,Lucini:2004my,Lucini:2004yh,Lucini:2003zr,Lucini:2002ku,Lucini:2012wq}
 is directed at the properties of pure gauge theory.
I know of two papers on meson spectroscopy:
Refs.~\cite{DelDebbio:2007wk,Bali:2008an}.
They are done at smaller values of ``large $N$,'' $N=2$, 3, 4, 6.
They are pretty standard lattice QCD spectroscopy calculations and reveal the N-independence
 of meson masses.
Narayanan, Neuberger, and collaborators
\cite{Narayanan:2005gh,Hietanen:2009tu} have measured masses and the pseudoscalar decay constant
from simulations at 
 much larger $N$  (exploiting
reduction, to simulate on smaller lattice sizes). 

But only Ref.~\cite{Jenkins:2009wv} discusses large-N expectations for baryons, and
it only makes comparisons to actual lattice data for $N=3$.
So it seemed like an appropriate time to look at a lattice calculation
of baryon spectroscopy at several values of $N$.

Baryons in large N seem to be fascinating
 objects, either viewed as many-quark states \cite{Witten:1979kh} or as topological objects
in effective theories of mesons\cite{Witten:1983tx,Adkins:1983ya}.
There is an enormous (continuum) literature about spectroscopy and matrix elements for large $N$
baryons.
Assorted  early references include
\cite{ Jenkins:1993zu,Dashen:1993jt,Dashen:1994qi,Jenkins:1995td,Dai:1995zg},
summarized in a review, Ref.~\cite{Manohar:1998xv}.
Perhaps a lattice study might reveal something interesting?

In fact, it does: general arguments  \cite{Adkins:1983ya,Jenkins:1993zu} state that
the mass of an $N$ color baryon of angular momentum $J$ should show a rotor spectrum:
\bee
 M(N,J) = NA + \frac{J(J+1)}{N}B
\label{eq:jsplit}
\ee
The formula applies to baryons made of an $SU(2)$ isospin doublet of equal mass
quarks. The parameters $A$ and $B$ depend on the quark mass and both should be
some ``typical hadron size,'' a few hundred MeV. The observation of Eq.~\ref{eq:jsplit}
in simulation data
is the main new result of this paper. Common expectations are that Eq.~\ref{eq:jsplit}
is only true for small values of $J$, because then the terms have meaning
as a good expansion in $1/N$.
However, I observe that it holds both for the bottom and the top of the multiplets 
where I did measurements.

The first response of the lattice simulator to a proposal to look at baryons at large $N$ is undoubtedly
negative:  The cost of an $N$-color simulation scales roughly like $N^3$, just from the cost of multiplying
$SU(N)$ matrices. Baryons are large objects so big volumes will be needed. Ordinary
($N=3$) baryon correlators are noisy and because baryon masses scale as $N$, baryons in higher $N$ are
probably noisier. And most of the literature involves some combination of small $1/N$ effects and
expectation values of operators ($\xi\sim \langle B|O|B\rangle$), which are already hard enough
for $N=3$ \cite{Wittig:2012ha}.
 Fortunately, large $N$ tests seem (mostly) not to require extrapolations to zero quark mass.

Only for odd $N$ are baryons fermions, of course, so
I am performing  simulations of $SU(N)$ gauge theories with $N=3$, 5, 7.

I used the quenched approximation for all $N$. That means that, truly, I am not
simulating QCD for any value of $N$. However, a
first study does not quite demand the same level of quality as a later one might.
I am interested in comparing simple observables in systems
 which differ only in $N$, but otherwise have the same
UV cutoff (lattice spacing), lattice action, and physical volume. This is easy to set up
in quenched approximation. 

Of course, the system which is simulated ought to have some connection to
the real world. It is easy to justify a quenched approximation for large $N$
(although one could argue that it remains an open question, whether such justification is correct).
Quenched simulations for $N=3$ are obsolete, but  the effects of dynamical fermions
on spectroscopy are really not all that large: away from the deep chiral limit they are
small, order ten per cent for simple observables (compare Fig.~1 in Ref.~\cite{Davies:2003ik}).

The difficulty with using the quenched approximation comes when one wishes to convert
a lattice number to MeV. The spectrum of quenched QCD
is simply different from the spectrum of ``real'' QCD. I will compare dimensionless
ratios of lattice quantities in my tests of scaling with $N$.
Any conversions to MeV I make are only qualitative statements.

One could take the position that one should test everything about large-$N$ QCD at once,
by doing simulations with dynamical fermions at several values of $N$. This just
postpones the matching problem: the spectrum of the different $N$ QCD's will be different.
But perhaps that is not the right point of view. The real question one asks when
comparing several values of $N$ is whether dimensionless ratios of masses show
some smooth behavior with $N$. And for that, one can begin simply, make observations,
and then ask how they change as one does ever more realistic simulations.

For all $N$'s,
I use  lattices of size $16^3\times 32$  points.
I match the bare couplings so that the lattice spacings, as measured by various observables from
the heavy quark potential, are the same, and then sweep in quark mass over similar ranges.
 This insures that finite volume and nonzero lattice spacing lattice artifacts
will be similar across the board.  
I set the common scale using the shorter version of the Sommer \cite{Sommer:1993ce} 
parameter, $r_1$, defined in terms of the force $F(r)$
so that $r^2 F(r)= -1.0$ at $r=r_1$. The real-world value is $r_1= 0.31$ fm \cite{Bazavov:2009bb},
and with it my lattice spacings would be about 0.08 fm.

Before going on, it's useful to set some definitions.
The 't Hooft coupling is
\bee
g^2N=\lambda
\ee
and the usual gauge coupling is thus
\bee
\beta=\frac{2N}{g^2} = \frac{2N^2}{\lambda}.
\ee
The combination $g^2C_F$, which appears in all perturbative calculations of renormalization factors, is
equal to $\lambda(1-1/N^2)/2$. For comparisons at fixed 't Hooft coupling, like the ones I present,
this suggests that differences which might be perturbative (lattice--to--continuum matching factors, 
for example) will scale up to corrections of order $1/N^2$.

The outline of the paper is as follows: The next section describes some technical problems I faced:
using fat links, gauge fixing, and the construction of baryon operators. Only the third topic might be
of interest to continuum physicists.
 Next, I present lattice results for the potential and for meson masses and simple matrix elements.
The potential measurements show the extent to which lattice spacings are matched. The mesonic
observables
illustrate the $N$ -- independence of masses and the $\sqrt{N}$ scaling of quantities like the
pseudoscalar decay constant. Finally, in Section \ref{sec:baryons},
 I  show some results for baryon spectra. The major one is the
presence of a rotor spectrum of excitations for two-flavor baryons.

\section{Technical details\label{sec:details}}
\subsection{Simulation techniques}
My simulations use a version of the publicly available package of the MILC collaboration~\cite{MILC},
 modified to generate gauge configurations
and quark propagators at arbitrary $N$. Prior to this project, it had been extensively 
used in studies of $N=2$, 3, 4 \cite{DeGrand:2011qd} .

The gauge action is the usual Wilson action.
Quenched simulations are performed using the standard mix of Brown - Woch  microcanonical
 over-relaxation \cite{Brown:1987rra}
and
Cabibbo - Marinari  heat bath updates \cite{Cabibbo:1982zn}, performed
on all $N(N-1)/2$ $SU(2)$ subgroups of the $SU(N)$ link variables.
It is known that simulations performed on $SU(2)$ subgroups suffer critical slowing down at larger
$N$, but the largest $N$ is only 7 and this problem did not appear.
Lattices are spaced 100 sweeps of the lattice apart, for later analysis.

The spectroscopy was intended to be a typical $SU(3)$ lattice QCD calculation writ large.
This involved an improved fermion action and extended sources for hadronic correlation functions.
To achieve these goals a number of small technical problems had to be solved. Many of them have been
encountered
in other large-$N$ studies, but perhaps my solutions are a bit different than what is found there,
 and might be worth reporting.

\subsection{$SU(N)$ fat links; gauge fixing}

My lattice fermions are clover fermions with normalized hypercubic (nHYP-link) smeared links
as their gauge connections\cite{Hasenfratz:2007rf}.
The clover coefficient is fixed at its tree level value,
$c_{SW}=1$. This particular discretization is known to work well, with small scaling violations,
in both ordinary QCD phenomenology and in beyond-Standard Model studies in $SU(2)$, $SU(3)$, $SU(4)$.
So the first technical problem is  the construction of the nHYP link for arbitrary $N$.
To describe the nHYP link in words, it is a local average
of gauge connections over the hypercubes surrounding the link, which smears out the gauge field
for the fermion.
The specific problem to be solved is that the
the fat link is the average of a set of paths which produces a sum of $SU(N)$ matrices, call it $\Sigma$.
The fat link $V_\mu$ is defined as
\bee
V_\mu = \frac{ \Sigma}{\sqrt{\Sigma^\dagger \Sigma}}
\label{eq:fat}
\ee
This is the matrix which maximizes $\Re \Tr V_\mu \Sigma^\dagger$.
The quantity $ Q^{-1/2}=1/\sqrt{\Sigma^\dagger \Sigma }$ is computed using
 the Cayley - Hamilton theorem as described in  Appendix B of Ref.~\cite{DeGrand:2012qa}.
This construction involves finding the eigenvalues of $Q$, which is done using a Jacobi algorithm.

As written, $V_\mu$ is an element of $U(N)$, not $SU(N)$. This is not an issue when it is used
for the fermions since all that we care about is that our action should be gauge invariant, and under a
 gauge transformation both the thin and fat links transform the same way.

State of the art spectroscopic calculations use extended sources as interpolating fields for hadrons.
In this work configuration of link variables is gauge fixed to lattice Coulomb gauge and the source
for the quark propagator is some spatially extended function.
So, we need to gauge fix our lattices to lattice Coulomb gauge, rotating the links to maximize
$\sum_x \sum_i \Re \Tr U_i(x)$, $i$ labeling the spatial directions. This is done by finding a matrix
$V$ at each site of the lattice $x$ which maximizes $\Re \Tr V \Sigma^\dagger$ where $\Sigma$ is
 the sum of forward going
$U_\mu(x)$'s emanating from $x$ minus the sum of $U^\dagger(x-\mu)$'s terminating on site $x$.

In the author's $SU(3)$ code, gauge fixing is done iteratively by sweeping through the lattice
 and, site by site, determining an optimal $V$ by maximizing 
 $\Re \Tr V \Sigma^\dagger$. As in the case of configuration generation, the relaxation is done
using the $SU(2)$ subgroups of $V$. Unfortunately,
for $N>4$ relaxing on the subgroups suffers critical slowing down and it becomes impossible
to carry out gauge fixing without performing an enormous number of iterations.

This is a variation of a problem which has been previously observed (and solved) by several
groups simulating large $N$ gauge theories, with $N>10$ \cite{Kiskis:2003rd,deForcrand:2005xr}.
 There the problem is in the update
step. Updating on $SU(2)$ subgroups produces a simulation algorithm with a long autocorrelation time,
which becomes longer with increasing $N$.
The solution is to perform over-relaxation on the full $SU(N)$ group, implementing an old idea of
Creutz \cite{Creutz:1987xi}. I do this a little differently than 
Refs.~\cite{Kiskis:2003rd,deForcrand:2005xr}. The maximizing $V$ is 
given by Eq,~\ref{eq:fat}, which I have already dealt
with while generating the fat links for the fermion 
action. The same application of the Cayley-Hamilton theorem gives $V$. 
As I already remarked, this $V$ is an 
element of $U(N)$, not $SU(N)$. Now one needs $SU(N)$ elements, so I must compute the determinant of $V$,
extracting its phase $\phi$ and performing an additional
multiplication by the diagonal matrix $\exp(-i\phi/N)$.
Doing this makes gauge fixing no more expensive for $SU(7)$ than for $SU(3)$.

\subsection{Baryon operators}

Mesonic states are constructed in the usual way, by sandwiching fermion propagators with
Dirac matrices. Baryons are a bit more complicated. In $SU(3)$ it is common to use
relativistic sources ($\epsilon_{abc}[u^aC\gamma_5d^b]u^c$, for example, for the proton).
I don't know a nice way to generalize this to $N>3$, so
I built baryon states using non-relativistic quark model states. So far, I have only considered
 $N_f=2$ flavors of quarks. In what follows, I will label the flavors as 
$u$ and $d$, although one can give them different masses -- in $SU(3)$
 the same operator can give the $p$ or $\Xi$, for example. 

In the MILC code, quark propagators are constructed in the Weyl ($\gamma_5$ diagonal) basis, 
and are rotated to $\gamma_0$ basis.
I keep the two ``large'' components of the propagator's source and sink spin indices to
be the fields which contract the quark model states.
There are two choices for projection (eigenvalues of $\gamma_0=\pm 1$ allow forward
 going in time and backward going two-component quark propagators). One projection
has the lightest positive parity state in a channel as the forward-going state, and
the lightest negative-parity state as the backward-going one. The situation is reversed
for the other $\gamma_0$ projector. In order to reduce noise, these two propagators are summed
(actually subtracted), configuration by configuration.
 This produces a single correlator extending halfway across the lattice,
which is a candidate for fitting to a single dying exponential.

A generic $N-$color baryon operator can be written as
\bee
\ket{B}= \epsilon^{abc\dots N} \sum_{\{s_j\}} C_{\{s_j\}} u^a_{s_1} u^b_{s_2} \dots d^N_{s_N}
     \ket{0} .
\label{eq:bar}
\ee
The $C$'s are an appropriate set of Clebsch-Gordon coefficients. The baryon propagator
is a multiple sum over source and sink colors and spin coefficients of products of $N$
quark propagators.
This creates the possibility a large number of terms. Fortunately, many are redundant.

I condense states as follows. (I suspect that this is quite inefficient and that
more  efficiency is almost certainly possible.)
I begin with states of definite $u$ and $d$ content. These states are eigenstates
of $I_3$, $J$, and $J_3$. I then anticommute fermion fields into a ``standard order,''
shown in Eq.~\ref{eq:bar}: moving from the left, spin-up $u$ quarks, spin-down $u$ quarks,
spin-up $d$ quarks, and spin-down $d$ quarks. The epsilon symbol absorbs the resulting minus signs. 

Then, two examples are the $I=J=J_Z=N/2$ state
\bee
\ket{B,I=J=N/2}= \epsilon^{abc\dots N}u_\uparrow^a \dots  u_\uparrow^N \ket{0}
\label{eq:delta}
\ee
and the $SU(3)$ proton operator
\bee
\ket{p,\uparrow} = \epsilon^{abc} \sqrt{\frac{2}{3}}(u_\uparrow^a u_\uparrow^b d_\downarrow^c
 - u_\uparrow^a u_\downarrow^b d_\uparrow^c)
         \ket{0}.
\ee

Wick's theorem says that the $n-$quark propagator itself is also a determinant:
\beea
\langle q_{i_1}(x) \bar q_{j_1}(y) &q_{i_2}(x)& \bar q_{j_2}(y) \dots q_{i_n}(x) \bar
q_{j_n}(y)\rangle \nonumber \\
& =& (-)^n \sum_{P(1,2\dots n)} 
 {\rm sign}(P)(D^{-1}_{i_1,j_{P_1}} (x-y) D^{-1}_{i_2,j_{P_2}}(x-y) \dots D^{-1}_{i_n,j_{P_n}}(x-y) )
\nonumber \\
\label{eq:barprop}
\eea
($P$ is a permutation of indices). The baryon propagator, then is a product of an up-quark determinant times
a down quark determinant, summed over all the color combinations of the individual up and down quarks.
Many terms are redundant in this product, as can be seen in an example, the $SU(3)$ proton:
 The contributions from the
$\epsilon^{abc} u_\uparrow^a u_\uparrow^b$ source terms are antisymmetric in both Eqs.~\ref{eq:bar}
and \ref{eq:barprop}.
This means that it is only necessary to keep one color ordering $(b>a$, for example)
for each pair of
colors $(a,b)$ in the baryon propagator, and each term can be reweighted by a factor of $2!$.
This generalizes straightforwardly so that a a term in any wave function with 
$N_\uparrow^u$ spin-up $u$ quarks,
$N_\downarrow^u$ spin-down $u$ quarks,
$N_\uparrow^d$ spin-up $d$ quarks,
$N_\downarrow^d$ spin-down $d$ quarks,
picks up a restricted color sum, a single color ordering for each individual spin-flavor label and a multiplicity
$N_\uparrow^u! N_\downarrow^u! N_\uparrow^d!  N_\downarrow^d!$. 

An extreme example of this pruning procedure is the
 the propagator for the $I=N/2,J=N/2$ state of Eq.~\ref{eq:delta}. It is
\bee
\Delta(x,y)=(N!)^2 \det M(x,y)
\ee
where $M$ is the $N\times N$ matrix of $u_\uparrow$ propagators from a source color to a sink one.

The lower the $J$, the more complicated are the wave functions.
By $SU(7)$, the $J=1/2$ states involve several hundred color combinations per spin
configuration. This is getting quite unwieldy. (Since the nonrelativistic
quark propagator is itself only a $(2N)\times (2N)$ matrix,
there must be more redundancy!) But for now, I go on, naively.

To summarize, flavor $SU(2)$ wave functions and propagators for the various states
are
\begin{itemize}
\item $I=J=N/2$: as already described, there is one determinant of an $N\times N$ matrix, for a cost $N^3$.
\item Analogs of $\Sigma^*$ and $\Xi^*$ states are
\bee
\ket{B,I=N/2-1,J=N/2} =\sqrt{(N-1)!}(u_\uparrow^a \dots  u_\uparrow^{N-1}s_\uparrow^N \ket{0}.
\ee
The baryon correlator is built of  $N^2$ terms, one for each $s$ color in the source and sink.
 Each is a determinant of
an $(N-1)\times (N-1)$ matrix -- the propagator of the up quarks.
\item  $I=J=N/2-1$: We have an $(N-1)\times(N-1)$ matrix of $u-$quark propagators and
$N+N(N-1)$ color terms in the interpolating field for a cost of roughly $N^{4+3}=N^7$:
\bee
\ket{B,I=J=N/2-1}= \sqrt{\frac{N-1}{N}} (u_\uparrow^a \dots  u_\uparrow^{N-1}
d_\downarrow^N - u_\uparrow^a \dots  u_\uparrow^{N-2} u_\downarrow^{N-1}d_\uparrow^N)
\ket{0}
\ee
\item  $I=J=N/2-2$: An $(N-2)\times(N-2)$ matrix of propagators. The last term in the
interpolating
field has about $N^3$ color possibilities, for a cost of $N^9$.
\beea
\ket{B,I=J=N/2-2} &\propto&  (u_\uparrow^a \dots u_\uparrow^{N-2} d_\downarrow^{N-1}d_\downarrow^{N} \nonumber \\
& & -2  u_\uparrow^a \dots u_\uparrow^{N-3} u_\downarrow^{N-2}d_\uparrow^{N-1}d_\downarrow^{N}
\nonumber \\
& & + u_\uparrow^a \dots u_\uparrow^{N-4}u_\downarrow^{N-3}u_\downarrow^{N-2}d_\uparrow^{N-1} d_\uparrow^N)
  \ket{0}\nonumber \\
\eea
\item  $I=J=N/2-3$: our lowest state if $N=7$; there are $N^{11}$ terms to evaluate.
\beea
\ket{B,I=J=N/2-3} &\propto&  (u_\uparrow^a \dots u_\uparrow^{N-3} d_\downarrow^{N-2}\dots d_\downarrow^{N} \nonumber \\
& & -3 u_\uparrow^a \dots u_\uparrow^{N-4}u_\downarrow^{N-3} d_\uparrow^{N-2}d_\downarrow^{N-1}d_\downarrow^{N} \nonumber \\
& & +3 u_\uparrow^a \dots u_\uparrow^{N-5} u_\downarrow^{N-4}  u_\downarrow^{N-3}   d_\uparrow^{N-2}
 d_\uparrow^{N-1}d_\downarrow^{N} \nonumber \\
& & -u_\uparrow^a \dots u_\uparrow^{N-6}u_\downarrow^{N-5} \dots u_\downarrow^{N-3}
d_\uparrow^{N-2}\dots d_\uparrow^{N}) 
  \ket{0}\nonumber \\
\eea
\end{itemize}

Still to be constructed are three flavor ($u$, $d$, $s$) states.

An obvious solution to the problem of increasing multiplicity
 is to prune states by doing an incomplete color sum.
This will collide with
 another big problem: $SU(N)$ baryon signals degrade as $N$ increases.
For now, the  only  way to fight this is to collect larger data sets.
Along the way, however, one can try to improve our signals by 
tactics such as averaging over the propagators with the same $J$ and different
$m_J$'s.
 (In practice, I combine
the $m_J$ and $-m_J$  propagators into a single correlator.) Another approach involves
forcing a fit of several correlators which couple to the same states
to a common mass.  Ultimately,
 a variational calculation along the lines of what is done for
excited state baryon spectroscopy might be necessary.

\section{Results for pure gauge and mesonic observables\label{sec:mesons}}

I present quenched results from three $N$'s, $N=3$, 5, 7. I have data sets of 80, 120, and
160 propagators, respectively
The bare gauge coupling  $\beta$ is tuned to match potentials
through the Sommer parameters $r_0$ or $r_1$. (Recall  $r_0^2 F(r_0)=-1.65$. 
$r_1$ is the shorter-distance version of the Sommer parameter, $r_1^2 F(r_1)= -1.00$.
In the real world, $r_1\sim 0.31$ fm.)

\begin{figure}
\begin{center}
\includegraphics[width=0.9\textwidth,clip]{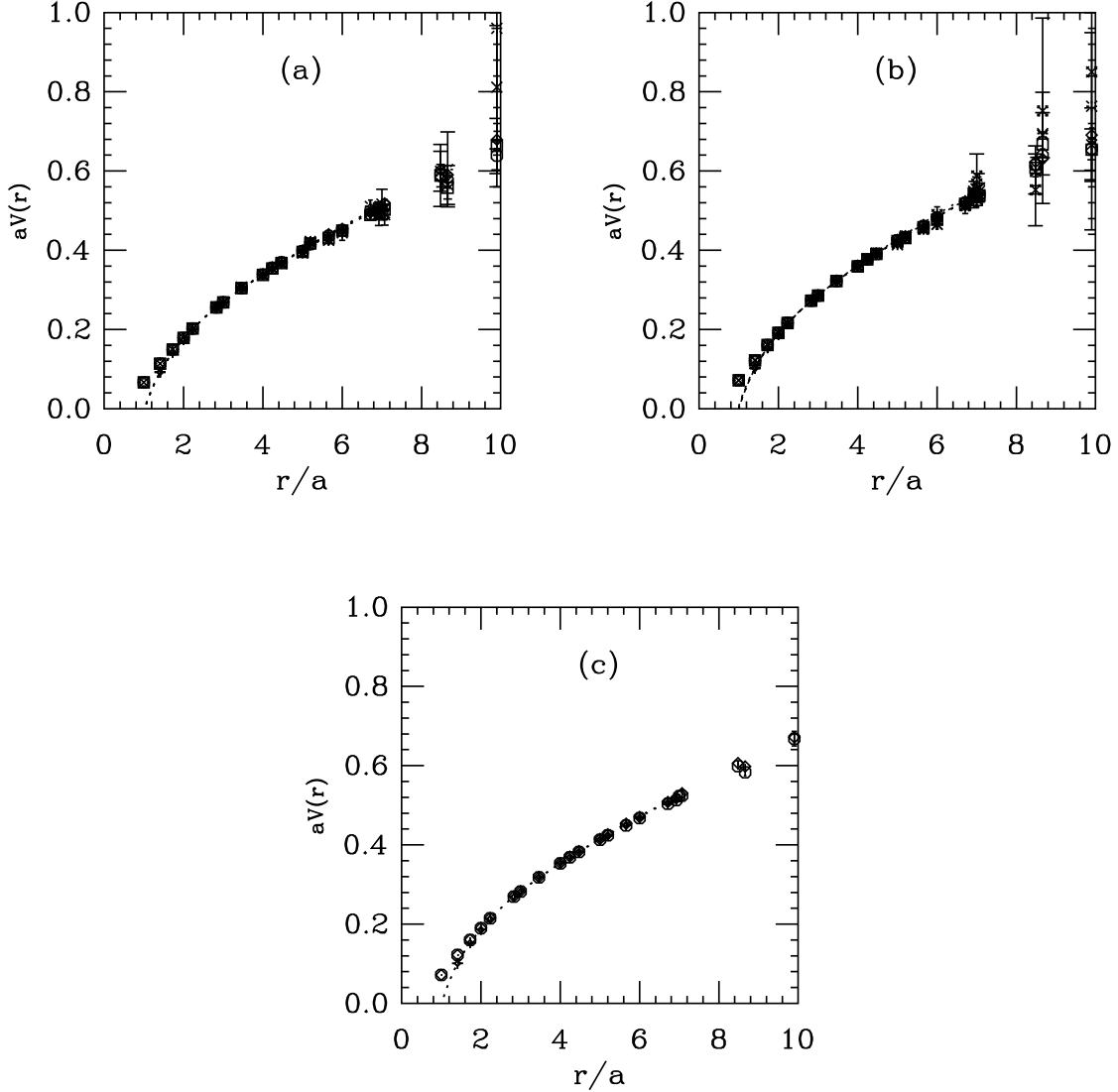}
\end{center}
\caption{Static potentials from our data sets:
(a) $SU(3)$ at $\beta=6.0175$, $16^3\times 32$ volume;
(b) $SU(5)$ at $\beta=17.5$,  $16^3\times 32$ volume;
(c) $SU(7)$ at $\beta=34.9$.  $12^3\times 32$ volume.
Effective mass fits for several values of $t$ are overlaid.
\label{fig:pot}}
\end{figure}

In the present simulations, $\beta(N=3)=6.0175$ (to match to previous work by Ref.~\cite{Bali:2008an}),
and then the $N=5$ and 7 couplings were tuned to match $r_1$. Couplings and derived quantities
are recorded in Table \ref{tab:vr}. The bare t Hooft couplings $\lambda$ turn out to be quite similar.
Wilson-action simulations for $N>3$ have a first order lattice-artifact transition, and I checked
that my simulations are above it.
Potentials for the three values of N are shown in  Fig.~\ref{fig:pot}. They seem satisfactorily
matched.

The combination $r_0\sqrt{\sigma}$ or $r_1\sqrt{\sigma}$ ($\sigma$ is the string tension)
 gives a dimensionless combination of the Sommer radius and the
string tension $\sigma$. The table shows that this quantity scales well.

\begin{figure}
\begin{center}
\includegraphics[width=0.9\textwidth,clip]{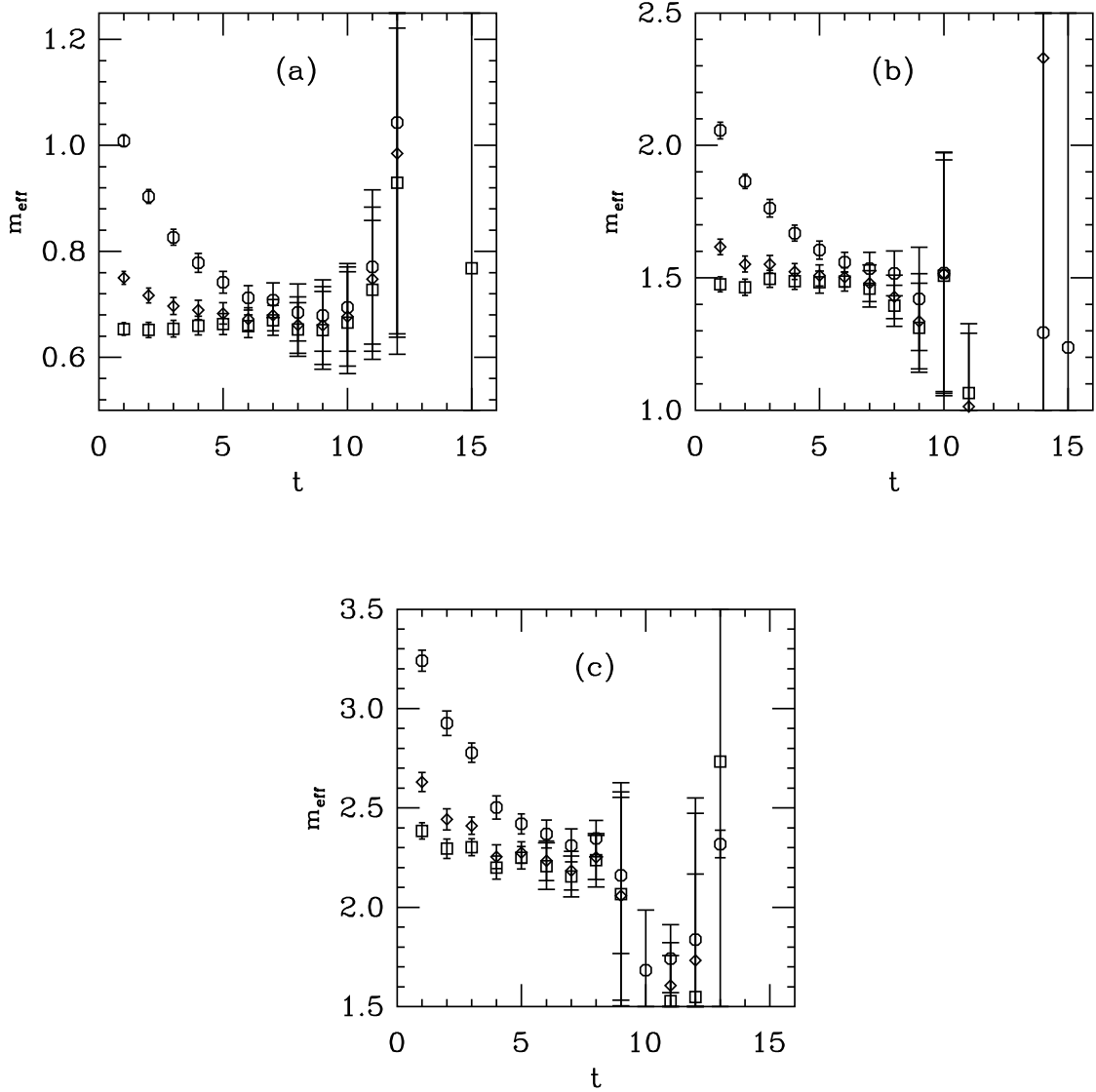}
\end{center}
\caption{Effective masses for the highest-spin baryon for different size sources,
labeled by octagons for $R_0=4$, diamonds for $R_0=6$, and squares, $R_0=8$.
(a) $SU(3)$, $\kappa=0.1265$;
(b) $SU(5)$ $12^3$ volume, $\kappa=0.1275$;
(c) $SU(7)$, $12^3$ volume, $\kappa=0.129$.
\label{fig:effdelt}}
\end{figure}

My spectroscopy is based on smeared Gaussian sources and $\vec p=0$ point sinks.
At each $N$ I collected sets for several different values of the width of $R_0$ for the source.
These correlation functions are not variational since the source and sink are different.
Thus, as $R_0$ is varied, the effective mass ($m_{eff}$ is defined through fitting correlators
at a single distance to a single exponential; with open boundary conditions for correlator $C(t)$,
$m_{eff} = \log C(t)/C(t+1)$) can approach its asymptotic value from above or below. I observe that
typically, as $R_0$ rises, the mixed Gaussian - point correlators make their approach from 
above for smaller
$R_0$, and from below for bigger $R_0$. An example of this behavior is shown in Fig.~\ref{fig:effdelt}.
(At large $t$, the signal deteriorates -- a characteristic feature quite familiar from many $SU(3)$
studies. The noise
 is enhanced by the small data sets -- 40 lattices -- used to make the figure.)
Then, rather than re-running the propagator code with yet
another source, I can combine pairs of sources to produce a flat $m_{eff}$ distribution using say
\bee
C(t) = C(R_0=6,t)+ fC(R_0=8,t).
\ee
For $SU(3)$, the optimal source is a linear combination of $R_0=4$ and 6 sources, favoring larger $R_0$
 as the
quark mass falls. For $SU(5)$ I mix sources with $R_0=6$ and 8.
For $SU(7)$ the $R_0=8$ source produced flat effective mass plots across my mass range, and I did not
do any source mixing.

Tables \ref{tab:su3}-\ref{tab:su7} contain the resulting spectroscopy. The values of the
masses are highly correlated because they come from the same underlying configurations.
Mass differences, which will be described below, are taken from jackknife averages of the data.

Now I turn to results for mesonic observables. The critical hopping parameter
 $\kappa_c$ is determined through the vanishing point for the Axial Ward Identity (AWI)
quark mass, defined as
\bee
\partial_t \sum_\bx \svev{A_0^a(\bx,t)\co^a} = 2m_q \sum_\bx \svev{ P^a(\bx,t)\co^a}.
\label{eq:AWI}
\ee
where the axial current $A_\mu^a=\bar \psi \gamma_\mu\gamma_5 (\tau^a/2)\psi$, the pseudoscalar 
density $P^a=\bar \psi \gamma_5 (\tau^a/2)\psi$, and $\co^a$ could
be any source. Here it is my Gaussian shell model source.

\begin{figure}
\begin{center}
\includegraphics[width=0.5\textwidth,clip]{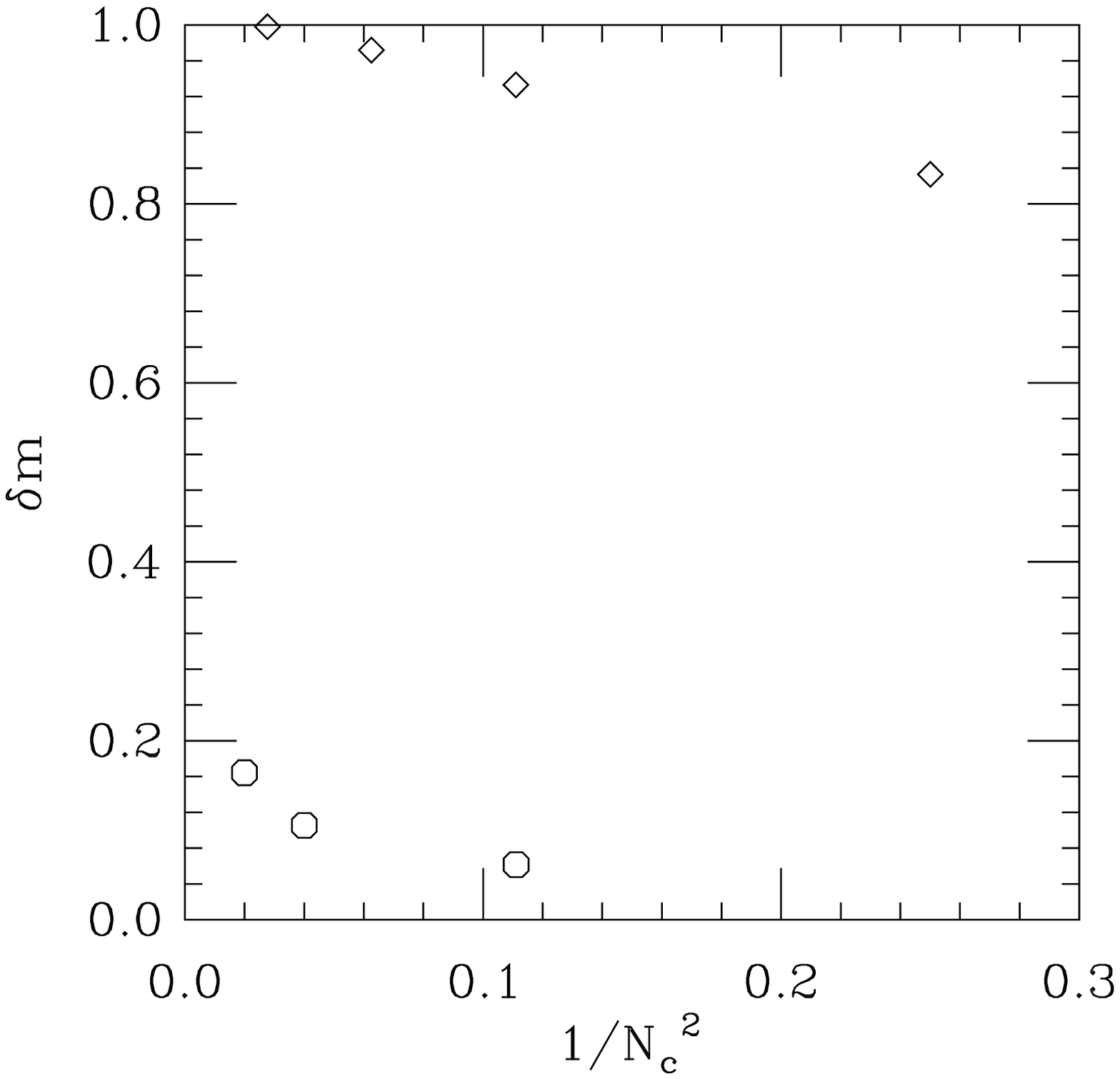}
\end{center}
\caption{Additive mass shift versus $1/N^2$. My data are shown as octagons. The diamonds are the pure
Wilson fermion data of Ref.~\cite{DelDebbio:2007wk}, just for comparison over a wider $N$ range.
\label{fig:dm1nc2}}
\end{figure}

The mass is shifted from its free value in the usual way. To compare to usual expectations, recall the relation between
$g^2C_F$ and $\lambda$: we expect that the mass shift
\bee
\delta m = \frac{1}{2\kappa_c}-4
\ee
to show $1/N^2$ variation. This I roughly see; compare Fig.~\ref{fig:dm1nc2}.
Just for comparison over a wider $N$ range, I show the older results of
Ref.~\cite{DelDebbio:2007wk}, done with unimproved Wilson fermions.

\begin{figure}
\begin{center}
\includegraphics[width=\textwidth,clip]{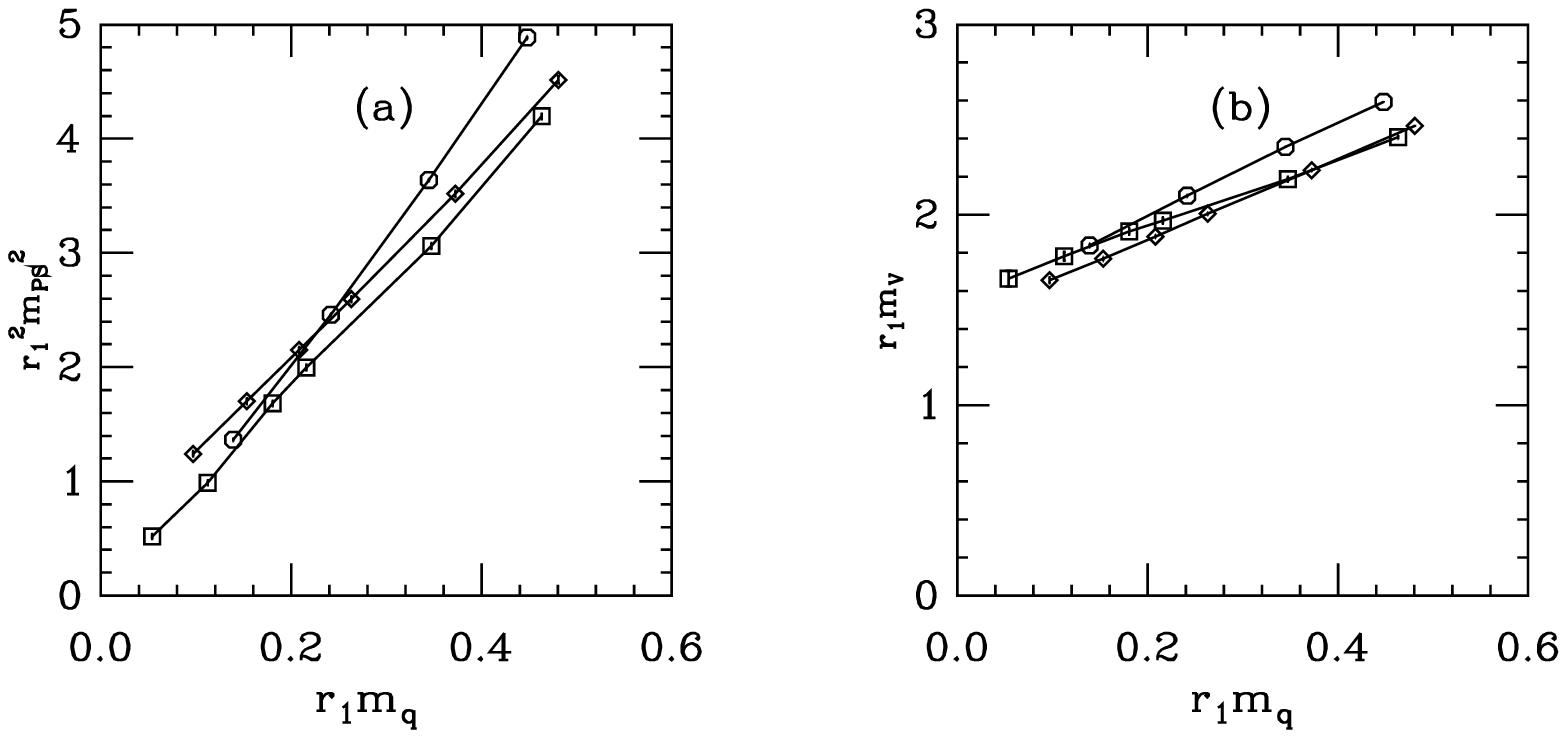}
\end{center}
\caption{Meson spectroscopy  in  units of $r_1$
from $N=3$, 5 and 7, plotted as squares, diamonds and octagons, respectively.
Panel (a) shows the squared pseudoscalar mass. 
Panel(b) shows the vector meson mass.
\label{fig:mesons}}
\end{figure}

\begin{figure}
\begin{center}
\includegraphics[width=0.6\textwidth,clip]{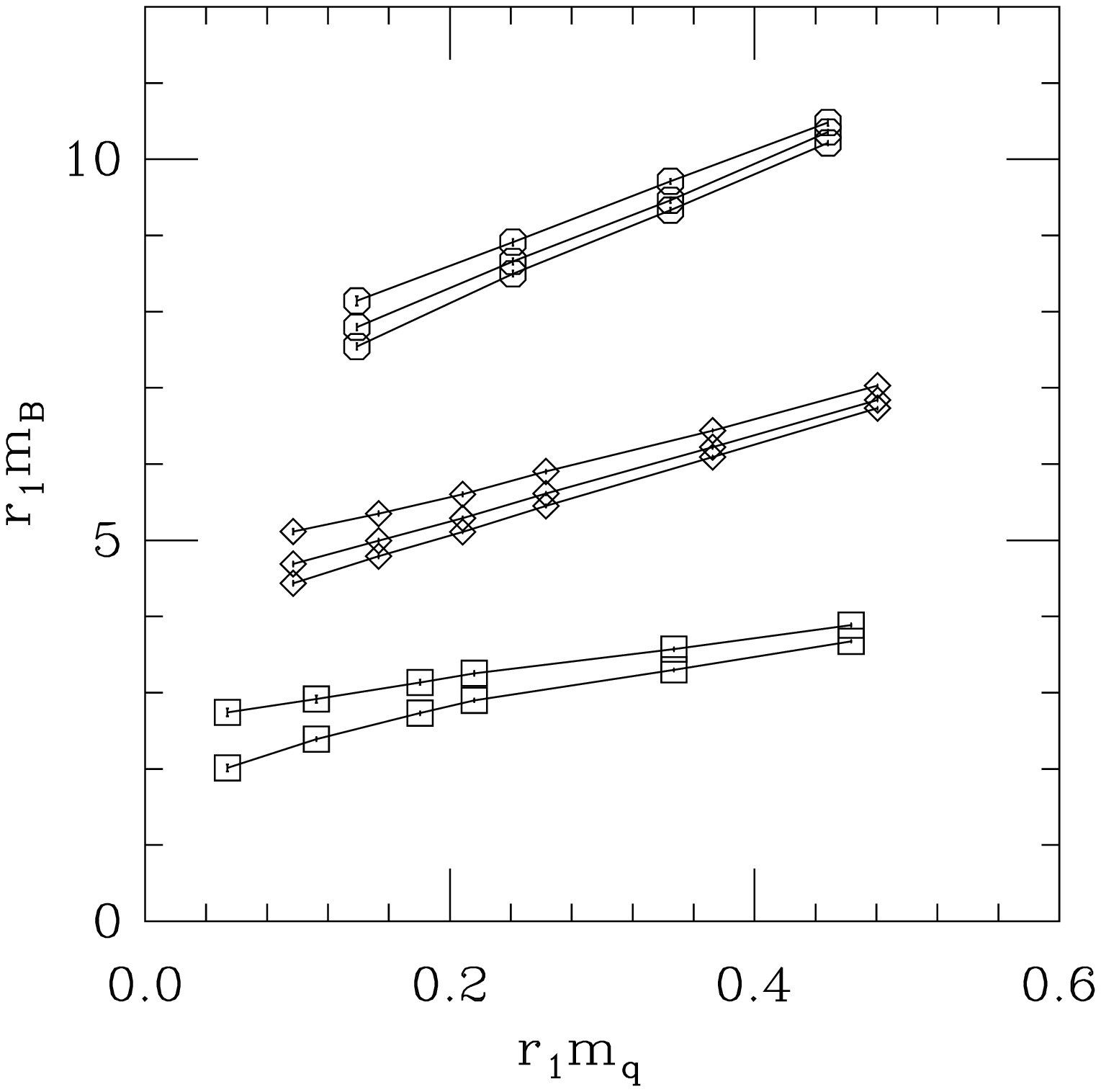}
\end{center}
\caption{Baryon spectroscopy  in units of $r_1$
from $N=3$, 5 and 7, plotted as squares, diamonds and octagons, respectively. For all $N$, higher $J$ lies higher in mass:
for $N=3$ and 5 I have all states from $J=1/2$ to $J=N/2$. For $N=7$ only the $J=7/2$, 5/2, and 3/2 states are shown.
\label{fig:baryons}}
\end{figure}

Next, we turn to spectroscopy, shown for mesons in
Fig.~\ref{fig:mesons} and baryons in Fig.~\ref{fig:baryons}.
Data are plotted in terms of  the AWI
quark mass, in units of $r_1$, to make the  $x$-axis  the same for all $N$.
 The near independence of meson masses on N (when expressed
in terms of a common variable) is apparent. Not so, for baryons!

One is tempted to compare the chiral limit of the vector meson mass as a function of $N$. A simple
linear fit to the data of Fig.~\ref{fig:mesons} gives $r_1 m_V=1.58(2)$, 1.44(1) and 1.52(1) for $N=3$,
5, 7. With a nominal $1/r_1=635$ MeV,   this is about 900 MeV to 1 GeV versus
770 MeV for the physical rho meson. One always has to be careful with 
quenched lattice results from a single
volume and
lattice spacing, but the number is completely sensible. It is close to the results
of Refs.~\cite{DelDebbio:2007wk,Bali:2008an}.

(Using completely different methodology plus a different choice of physical input to set the lattice
scale, the authors of Ref.~\cite{Hietanen:2009tu} have a mass which is from thirty per cent to a
factor of two
higher than this. The difference between their result and the low-N ones will probably
only be resolved by removing the many differences in methodology one at at time.)

In what follows, I will replace the quark mass as the independent variable by the square of the ratio
of the pseudoscalar to vector meson mass. The quark mass is scheme dependent. Of course, nothing I say is affected by
this choice.

Vacuum-to-meson matrix elements are expected to scale as $\sqrt{N}$.
 I have looked at the pseudoscalar, vector meson,
and axial vector meson decay constants. To be explicit, they are defined as follows:
\bee
\langle 0| \bar u \gamma_0 \gamma_5 d |\pi\rangle = m_\pi f_\pi
\ee
(so $f_\pi\sim 132$ MeV)
while the vector meson decay constant of state $V$ is defined as
\bee
\langle 0| \bar u \gamma_i d  | V\rangle = m_V^2 f_V \epsilon_i
\ee
and the axial vector meson decay constant of state $A$ is
\bee
\langle 0|  \bar u \gamma_i \gamma_5 d  |A \rangle = m_A^2 f_A \epsilon_i.
\ee
$\epsilon_i$ is a unit polarization vector. The lattice quantities $f^L$ are converted to
 continuum convention by 
$f=f^L(1-(3\kappa)/(4\kappa_c))$. I have left out the lattice-to-continuum Z-factor. With
 nHYP clover fermions,
it is a few percent away from unity.

\begin{figure}
\begin{center}
\includegraphics[width=0.6\textwidth,clip]{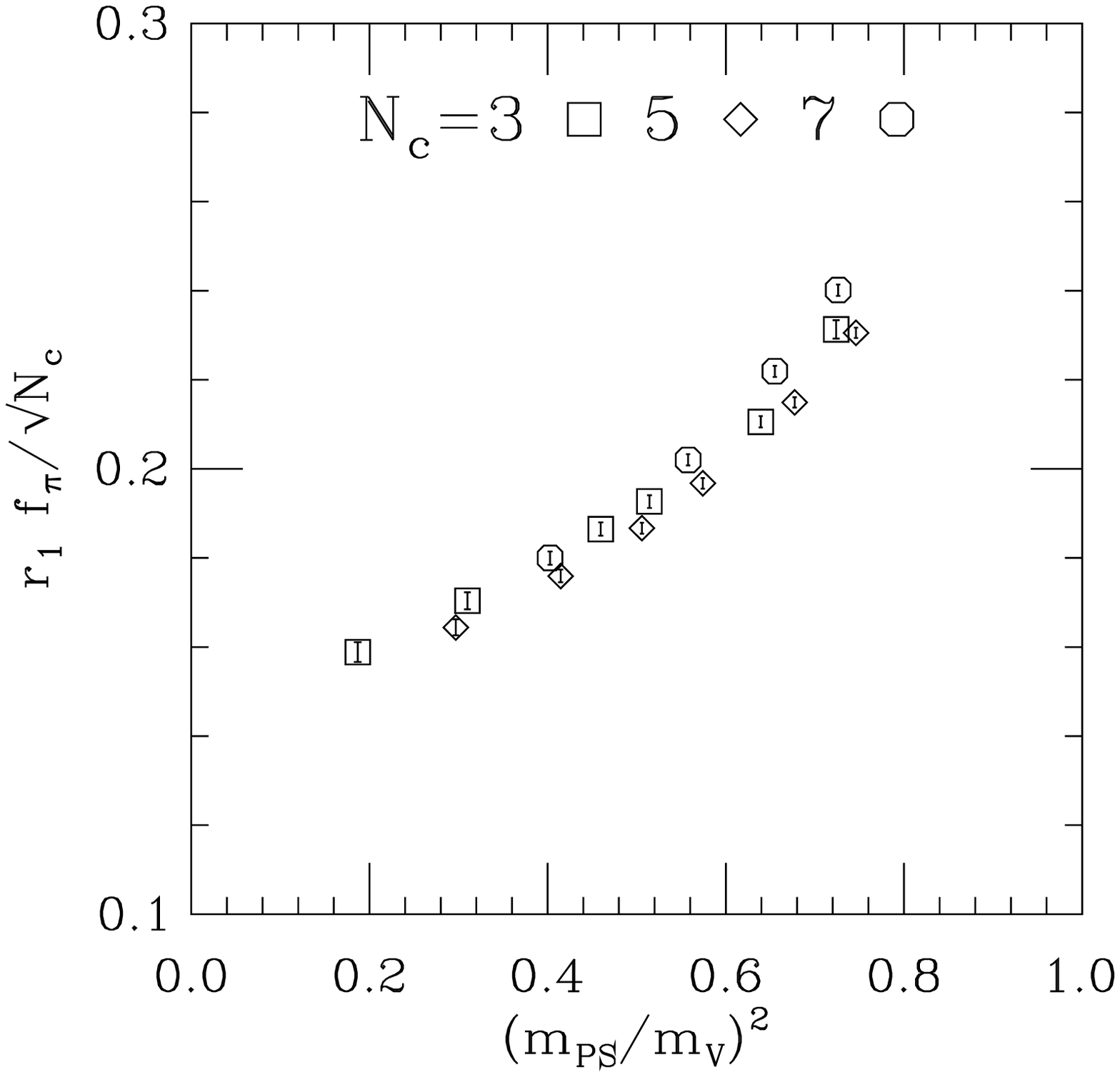}
\end{center}
\caption{Pseudoscalar decay constant, matched from the different simulations
using the $r_1$ parameter from the potential and rescaled by $1/\sqrt{N}$.
Squares, $N_c=3$;
diamonds, $N_c=5$,
octagons, $N_c=7$.
\label{fig:fpi}}
\end{figure}

The pseudoscalar decay constant is shown in Fig.~\ref{fig:fpi}. I show the dimensionless 
combination $r_1 f_{PS}/\sqrt{N}$. It's nice to see the $\sqrt{N}$ scaling.
Naive linear extrapolations give $r_1 f_\pi/\sqrt{N}=0.154(2)$, 0.151(1) and 0.154(2) for $N=3$, 5, 7.
The real world value is $1/\sqrt{3}\times 0.31$ fm $\times 132$ MeV $=0.12$, so the quenched decay constant at this
lattice spacing
is coming in about 15 per cent high. Experts know that this kind of extrapolation is far too naive,
to say nothing about comparing quenched QCD to the real world. 
Nevertheless, the answer is not absurd.
Ref.~\cite{Davies:2003ik} shows a figure of with the ratio of quenched $f_\pi$ to its experimental
result about ten per cent high.

$\sqrt{N}$ scaling for the pseudoscalar decay constant was first observed by the authors of 
Ref.~\cite{Narayanan:2005gh}. Their numerical result for the decay constant,
 translated to $SU(3)$, is also high, but by 40 per cent. 
Detecting the origin of this difference will probably again require detailed numerical work.

\begin{figure}
\begin{center}
\includegraphics[width=\textwidth,clip]{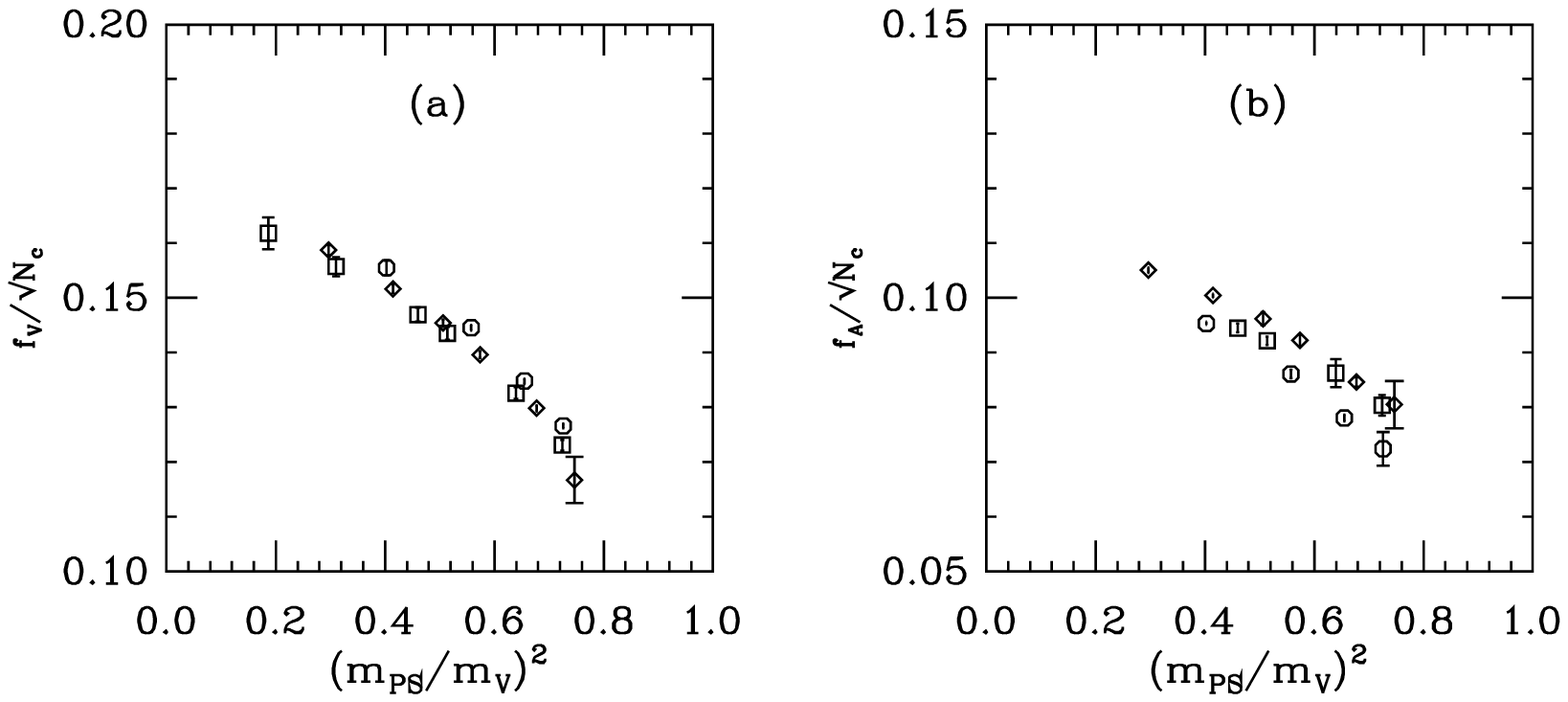}
\end{center}
\caption{Vector meson and axial vector meson decay constants,  rescaled by $1/\sqrt{N}$.
Squares, $N_c=3$;
diamonds, $N_c=5$,
octagons, $N_c=7$.
\label{fig:fva1}}
\end{figure}

 The same comparison 
is shown for $f_V$, and $f_{a_1}$ in Fig.~\ref{fig:fva1}. The $a_1$ decay constant is quite 
noisy at small
quark mass and I omit those results as untrustworthy.
 The $\sqrt{N}$ scaling rule works  well here, too.

\section{Results for baryons -- the rotor spectrum\label{sec:baryons}}

\begin{figure}
\begin{center}
\includegraphics[width=0.9\textwidth,clip]{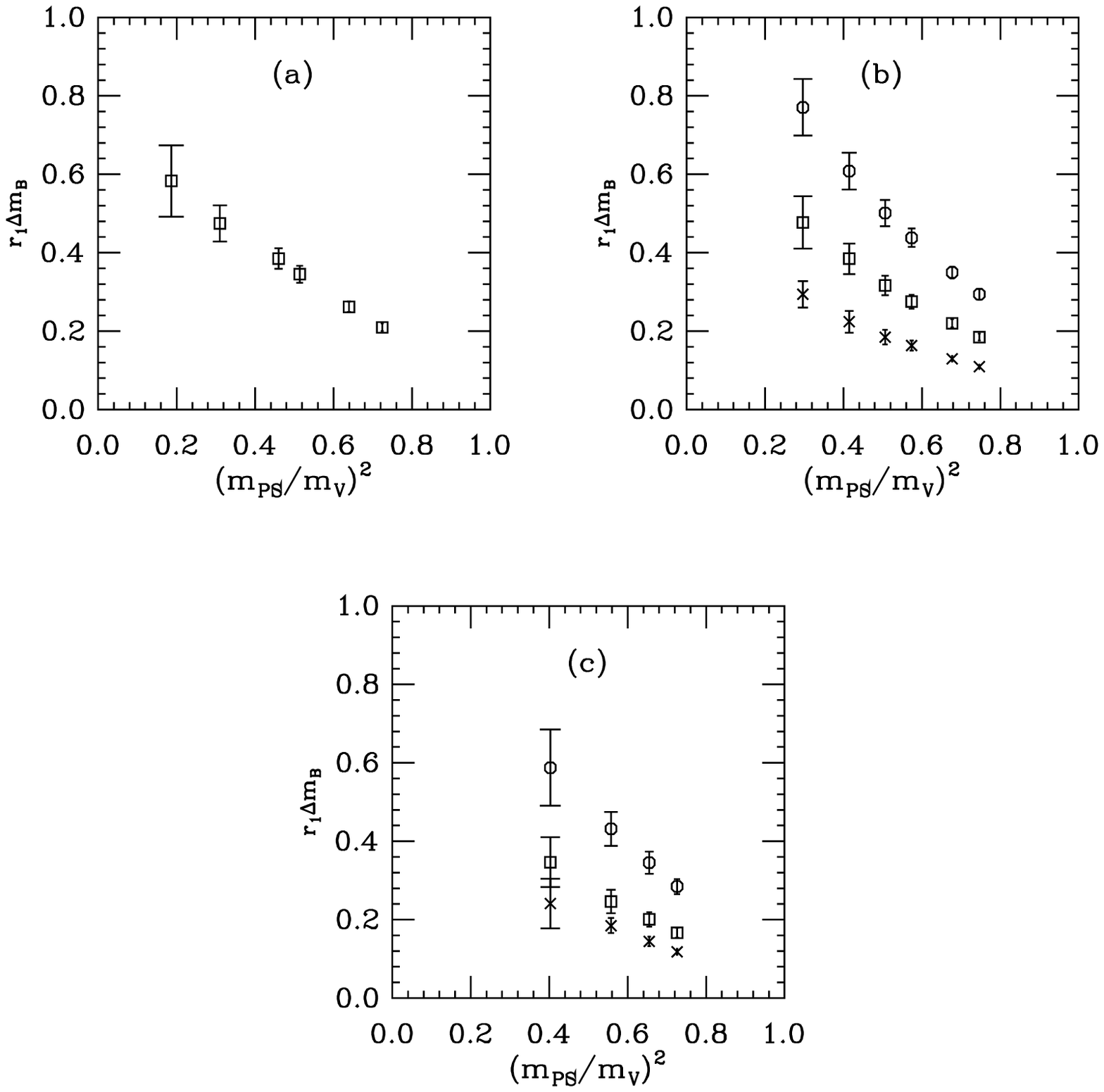}
\end{center}
\caption{(a) Delta-proton ($J=3/2-1/2$) mass splitting versus quark mass, in quenched $SU(3)$.
(b) $SU(5)$ mass splittings: octagons for $J=5/2-1/2$, squares for $J=5/2-3/2$,
and crosses, for $J=3/2-1/2$. In large-N, Eq.~\protect{\ref{eq:jsplit}} says that
 the $SU(5)$ $J=5/2-3/2$ splitting is supposed to
be equal to the $SU(3)$ $J=3/2-1/2$ mass splitting, and also to the $SU(7)$ $J=7/2-5/2$
splitting.
(c) $SU(7)$ mass splittings: octagons for $J=7/2-3/2$, squares for $J=7/2-5/2$,
and crosses, for $J=5/2-3/2$.
\label{fig:bardiff}}
\end{figure}

My baryon data in Fig.~\ref{fig:baryons} shows masses which increase roughly linearly in $N$.
All data shows that for the baryons, higher $J$ does lie higher in energy.
 This is no surprise,
so let's look deeper.  Fig.~\ref{fig:bardiff} shows the
 splitting between the various members of
each multiplet. It is extracted using a jackknife average of differences of the two baryon masses.

\begin{figure}
\begin{center}
\includegraphics[width=\textwidth,clip]{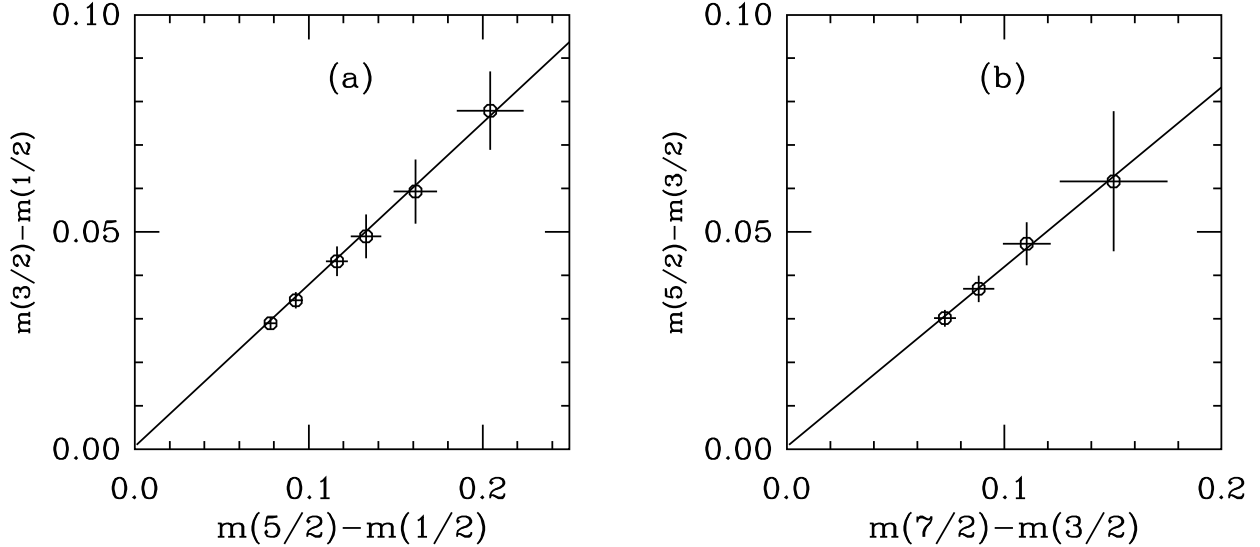}
\end{center}
\caption{Mass differences in the $SU(5)$ and $SU(7)$ multiplets, panels (a) and (b) respectively.
For $SU(5)$ I show a line whose  slope is the
 ratio $\Delta(5/2,3/2,1/2)$ (recall Eq.~\protect{\ref{eq:dj}}). The line has slope
3/8. For $SU(7)$ I show a line whose slope is  $\Delta(7/2,5/2,3/2) = 5/12$. 
\label{fig:dmdm}}
\end{figure}

I now demonstrate that the masses in  Fig.~\ref{fig:bardiff} form a rotor spectrum.
First, we can test the numerator of the rotor term of Eq.~\ref{eq:jsplit} $N$ by $N$.
This is done by looking at the ratio differences
\bee
\Delta(J_1,J_2,J_3)= \frac{M(N,J_2)-M(J_3)}{M(N,J_1)-M(N,J_3)},
\label{eq:dj}
\ee
for which  the constants ($A$, $B$) cancel. The result is shown in Fig.~\ref{fig:dmdm}. I just plot
one mass difference as a function of the other one and compare the data to a straight line
of zero intercept whose slope is given by Eq.~\ref{eq:dj}.
The rotor spectrum is confirmed for all the members of the $N=5$ and 7 multiplets I observe.

\begin{figure}
\begin{center}
\includegraphics[width=\textwidth,clip]{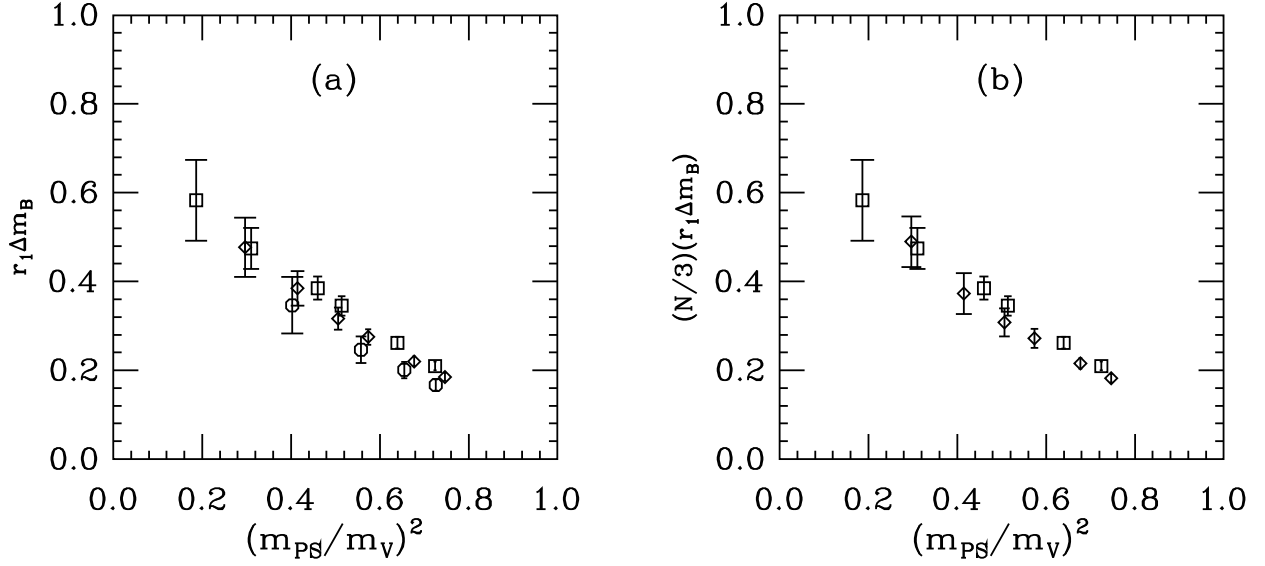}
\end{center}
\caption{
Exposing the $B$ term of Eq.~\protect{\ref{eq:jsplit}}:
(a) The $J=N/2$ vs $J=N/2-1$ mass difference in the $SU(3)$, $SU(5)$, and $SU(7)$ multiplets, 
shown respectively
as squares, diamonds, and octagons.
(b) $N/3$ times the $J=3/2-1/2$ mass differences in the $SU(3)$ and $SU(5)$ multiplets, shown
 respectively
as squares and diamonds.
\label{fig:Bterm}}
\end{figure}

Second, one can look
 at  the $J=3/2-1/2$ splitting, and check the $N$ dependence for the
 bottom of the multiplets:
\bee
M(N,3/2)=M(N,1/2)= \frac{3B}{N}
\label{eq:diff31}
\ee
Given the states I have recorded, this can only be done for $N=3$ and 5. Rescaling the
mass difference by $N/3$ exposes $B$.
The result is shown in Fig.~\ref{fig:Bterm}(a). This is quite promising: the $N=3$ and 5 data coincide.

Next, we can look at the top of the multiplet. Eq.~\ref{eq:jsplit} gives us
 a rescaled Land\'e interval rule: it
 says that the splitting between the $J=N/2$ and  $J=N/2-1$ states is a constant,
$B$, independent of
$J$.
Fig.~\ref{fig:Bterm}(b) shows this difference. It and
 panel (a)  share the common $N=3$ points, but the other points
are different. The envelope of the curve is $B(m_q)$.

\begin{figure}
\begin{center}
\includegraphics[width=\textwidth,clip]{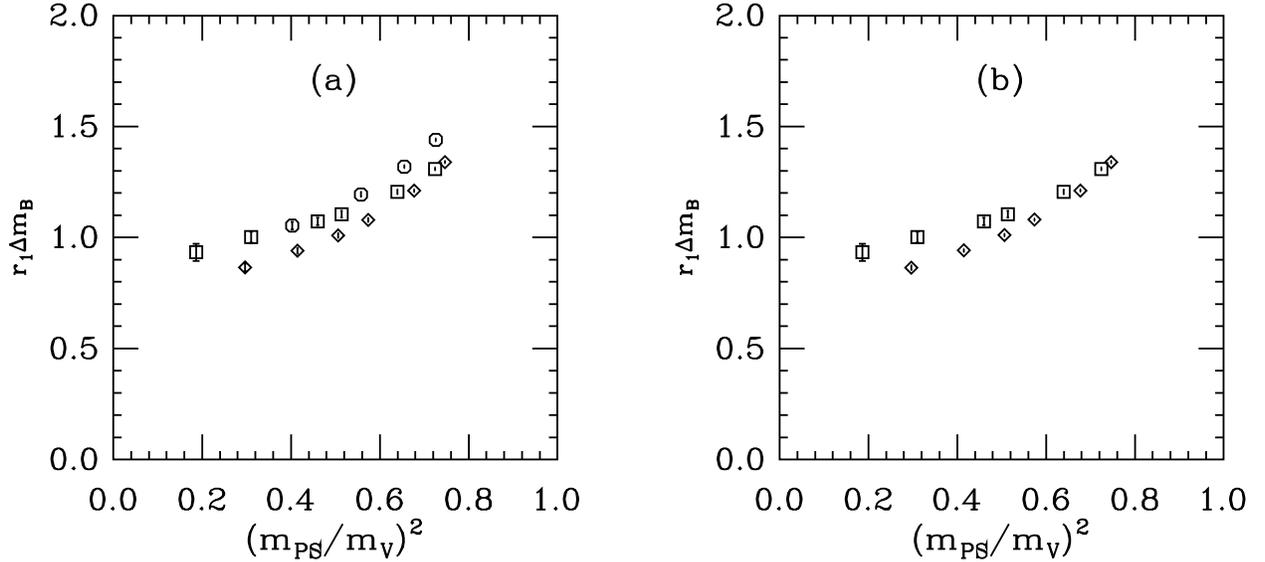}
\end{center}
\caption{
Exposing the $A$ term of Eq.~\protect{\ref{eq:jsplit}}:
(a) The $J=N/2$ vs $J=N/2-1$ mass difference of Eq.~\protect{\ref{eq:topA}}
in the $SU(3)$, $SU(5)$, and $SU(7)$ multiplets, 
shown respectively
as squares, diamonds, and octagons.
(b) Eq.~\protect{\ref{eq:botA}} for the $SU(3)$ and $SU(5)$ multiplets, shown
 respectively
as squares and diamonds.
\label{fig:Aterm}}
\end{figure}

By design, these differences ignore the $A$ term in  Eq.~\ref{eq:jsplit}. To get it, we can look
at the top of the multiplets,
\bee
A=\frac{N+2}{4N}M(N,J=N/2) - \frac{N-2}{4N}M(N,J=N/2-1)
\label{eq:topA}
\ee
or the bottom of the multiplets,
\bee
A=\frac{5}{4N}M(N,J=N/2) - \frac{1}{4N}M(N,J=N/2-1).
\label{eq:botA}
\ee
Fig.~\ref{fig:Aterm} shows these two mass formulas. Again, they behave consistently.

Figs.~\ref{fig:Bterm} and \ref{fig:Aterm} show that the $A$ and $B$ coefficients
in Eq.~\ref{eq:jsplit} have typical hadronic sizes. Inserting a nominal lattice spacing, $1/a\sim
2100$ MeV, we see that $A=400$ MeV at small quark mass and is an increasing function of quark mass.
In quark model language, $A$ is the constituent quark mass, and its value and dependence on quark mass
are both quite reasonable.
$B=300$ MeV and falls with energy. Again this is a  typical hadronic scale.

These values address an old conundrum of large-$N$ phenomenology:
the mass difference of the nucleon and $\Delta$ is supposed to be order $\Lambda_{QCD}/N$.
It is measured to be about 300 MeV, which is itself order $\Lambda_{QCD}$.
The plots of mass differences show that the nucleon-$\Delta$ mass splitting is indeed well-described
by large-$N$ QCD. Having data at several $N$'s as well as at several quark masses makes this
result appear immediately.
 
\section{Conclusions}

The qualitative features of
large - $N$  QCD phenomenology are easy to observe.
The shape of the potential (characterized by the scaling
combination $r_1\sqrt{\sigma}$) is $N$ - independent. 
Meson masses show little $N$ dependence -- $m_H r_1$ is $N$-independent.
Vacuum  - to -hadron matrix elements scale as $\sqrt{N}$.
Two-flavor baryons show a prominent rotor spectrum. This seems to be true for both the bottom
 of the spectrum (low $J$)
and the top ($J=N$ states).

 Naively, one might expect that the wave function of the baryon would be $N$ - independent.
Lattice simulations do not generally directly reveal wave function information, but one
 might expect that the same interpolating
fields might behave the same for different $N$'s. That does not seem to be the 
case; larger $N$ seems to prefer
larger Gaussian trial wave functions. But perhaps even $N=7$ is not such large
 $N$ and one should push farther.

Readers who do lattice simulations can undoubtedly list many 
obvious extensions of this project: smaller lattice
spacing, to do an honest extrapolation to the continuum limit,
 bigger volumes to check that the answers
 are trustworthy, smaller quark masses, and bigger $N$'s
are obvious technical improvements.
Giving the two fermions different masses, or better yet, considering flavor $SU(3)$,
would allow more tests of large-$N$ spectroscopy. 
One might really want to test whether dynamical fermions become less important at
large $N$.
 Writing the
code to simulate $SU(N)$ fundamental fermions is straightforward; running it might be costly,
and seeing the expected small differences shrink as $N$ rises would be even more costly.

The continuum literature of large - N baryons is more than thirty years old, 
and I clearly have only scratched its surface in this paper. Several simple ingredients 
were useful:
having data at several $N$'s, at several $J$'s for each $N$, and having data at many 
quark masses.
 Most of the continuum literature 
I have read restricts itself to statements
about $J=1/2$ and 3/2 -- presumably because that is all that exists in experimental 
data. Predictions for  any $J$ can challenge
the lattice and would be candidates for future studies.

\begin{table}
\begin{tabular}{c c c c}
\hline
                 & $SU(3)$   & $SU(5)$   &   $SU(7)$ \\
\hline
$\beta$          & 6.0175  & 17.5 & 34.9 \\
$\lambda=g^2N$   & 2.99    & 2.86 & 2.81 \\
$r_0/a$          & 5.36(6) & 5.20(5) & 5.41(3) \\
$r_1/a$          & 3.90(3) & 3.77(3) & 3.91(2) \\
$r_0\sqrt{\sigma}$ & 1.175(3) & 1.172(3) & 1.167(2) \\
$r_1\sqrt{\sigma}$ & 0.856(5) & 0.850(4) & 0.845(2) \\
 \hline
 \end{tabular}
\caption{Bare parameters and observables from potentials.
\label{tab:vr}}
\end{table}

\begin{table}
\begin{tabular}{c c c c c c }
\hline
$\kappa$  & $am_q$ & $am_{PS}$ & $am_V$  & $am_B(J=3/2)$  & $am_B(J=1/2)$ \\
\hline
0.1230 & 0.119 &  0.525(2)   & 0.617(3)  & 0.996(6)   & 0.942(5)  \\
0.1240 & 0.089 &  0.449(2)   & 0.561(3)  & 0.915(7)   & 0.846(5)  \\
0.1250 & 0.055 &  0.362(2)   & 0.505(4)  & 0.835(8)   & 0.744(6)  \\
0.1253 & 0.046 &  0.333(2)   & 0.491(5)  & 0.804(9)   & 0.700(6)  \\
0.1260 & 0.029 &  0.255(3)   & 0.457(6)  & 0.747(12)   & 0.613(7)  \\
0.1265 & 0.014 &  0.184(4)   & 0.427(10)  & 0.703(13)   & 0.516(12)  \\
 \hline
 \end{tabular}
\caption{Spectra from $SU(3)$ simulations.
\label{tab:su3}}
\end{table}

\begin{table}
\begin{tabular}{c c c c c c c}
\hline
$\kappa$  & $am_q$ & $am_{PS}$ & $am_V$  & $am_B(J=5/2)$  & $am_B(J=3/2)$  & $am_B(J=1/2)$ \\
\hline
0.1240 & 0.127 &  0.565(1)   & 0.655(1)  & 1.864(6)   & 1.814(5)  & 1.786(5)  \\
0.1250 & 0.099 &  0.488(1)   & 0.593(1)  & 1.708(6)   & 1.650(6)  & 1.616(5)  \\
0.1260 & 0.070 &  0.403(1)   & 0.532(2)  & 1.566(7)   & 1.488(6)  & 1.446(6)  \\
0.1265 & 0.055 &  0.356(1)   & 0.500(2)  & 1.486(7)   & 1.404(6)  & 1.356(6)  \\
0.1270 & 0.041 &  0.302(2)   & 0.469(3)  & 1.419(8)   & 1.325(7)  & 1.270(7)  \\
0.1275 & 0.026 &  0.240(2)   & 0.440(4)  & 1.357(10)   & 1.243(9)  & 1.177(8)  \\
 \hline
 \end{tabular}
\caption{Spectra from $SU(5)$ simulations.
\label{tab:su5}}
\end{table}

\begin{table}
\begin{tabular}{c c c c c c c}
\hline
$\kappa$  & $am_q$ & $am_{PS}$ & $am_V$  & $am_B(J=7/2)$  & $am_B(J=5/2)$  & $am_B(J=3/2)$ \\
\hline
0.1260 & 0.115 &  0.565(1)   & 0.663(1)  & 2.681(11)   & 2.649(8)  & 2.612(8)  \\
0.1270 & 0.088 &  0.488(1)   & 0.603(1)  & 2.483(9)   & 2.420(10)  & 2.386(9)  \\
0.1280 & 0.062 &  0.401(1)   & 0.537(2)  & 2.279(11)   & 2.215(11)  & 2.174(10)  \\
0.1290 & 0.036 &  0.299(1)   & 0.471(3)  & 2.082(15)   & 1.994(12)  & 1.929(11)  \\
 \hline
 \end{tabular}
\caption{Spectra from $SU(7)$ simulations.
\label{tab:su7}}
\end{table}

\begin{acknowledgments}
This project began as a direct result of my participation in the
Workshop on Large-N Gauge Theories at the Galileo Galilei Institute in Florence in May 2011.
I thank 
C.~DeTar,
R.~Lebed,
A.~Manohar,
H.~Neuberger,
and
M.~Teper
for discussions about this subject.
The modification of the MILC code to arbitrary number of colors was done with
Y.~Shamir and B.~Svetitsky.
This work was supported in part by the U.~S. Department of Energy.

\end{acknowledgments}



\begin{thebibliography}{99}

\bibitem{'tHooft:1973jz} 
  G.~'t Hooft,
  Nucl.\ Phys.\ B {\bf 72}, 461 (1974).

\bibitem{'tHooft:1974hx} 
  G.~'t Hooft,
  Nucl.\ Phys.\ B {\bf 75}, 461 (1974).


\bibitem{Lucini:2005vg}
  B.~Lucini, M.~Teper, U.~Wenger,
  JHEP {\bf 0502}, 033 (2005).
  [hep-lat/0502003].

\bibitem{Lucini:2004my}
  B.~Lucini, M.~Teper, U.~Wenger,
  JHEP {\bf 0406}, 012 (2004).
  [hep-lat/0404008].

\bibitem{Lucini:2004yh}
  B.~Lucini, M.~Teper, U.~Wenger,
  Nucl.\ Phys.\  {\bf B715}, 461-482 (2005).
  [hep-lat/0401028].


\bibitem{Lucini:2003zr}
  B.~Lucini, M.~Teper, U.~Wenger,
  JHEP {\bf 0401}, 061 (2004).
  [hep-lat/0307017].

\bibitem{Lucini:2002ku}
  B.~Lucini, M.~Teper, U.~Wenger,
  Phys.\ Lett.\  {\bf B545}, 197-206 (2002).
  [hep-lat/0206029].

\bibitem{Lucini:2012wq} 
  B.~Lucini, A.~Rago and E.~Rinaldi,
  arXiv:1202.6684 [hep-lat].



\bibitem{DelDebbio:2007wk}
  L.~Del Debbio, B.~Lucini, A.~Patella, C.~Pica,
  JHEP {\bf 0803}, 062 (2008).
  [arXiv:0712.3036 [hep-th]].

\bibitem{Bali:2008an}
  G.~S.~Bali, F.~Bursa,
  JHEP {\bf 0809}, 110 (2008).
  [arXiv:0806.2278 [hep-lat]].





\bibitem{Narayanan:2005gh}
  R.~Narayanan and H.~Neuberger,
  Phys.\ Lett.\  B {\bf 616}, 76 (2005)
  [arXiv:hep-lat/0503033].

\bibitem{Hietanen:2009tu}
  A.~Hietanen, R.~Narayanan, R.~Patel and C.~Prays,
  Phys.\ Lett.\  B {\bf 674}, 80 (2009)
  [arXiv:0901.3752 [hep-lat]].


\bibitem{Jenkins:2009wv}
  E.~E.~Jenkins, A.~V.~Manohar, J.~W.~Negele, A.~Walker-Loud,
  Phys.\ Rev.\  {\bf D81}, 014502 (2010).
  [arXiv:0907.0529 [hep-lat]].



\bibitem{Witten:1979kh}
  E.~Witten,
  Nucl.\ Phys.\  {\bf B160}, 57 (1979).


\bibitem{Witten:1983tx} 
  E.~Witten,
  Nucl.\ Phys.\ B {\bf 223}, 433 (1983).



\bibitem{Adkins:1983ya} 
  G.~S.~Adkins, C.~R.~Nappi and E.~Witten,
  Nucl.\ Phys.\ B {\bf 228}, 552 (1983).


\bibitem{Jenkins:1993zu}
  E.~E.~Jenkins,
  Phys.\ Lett.\  {\bf B315}, 441-446 (1993).
  [hep-ph/9307244].


\bibitem{Dashen:1993jt}
  R.~F.~Dashen, E.~E.~Jenkins, A.~V.~Manohar,
  Phys.\ Rev.\  {\bf D49}, 4713 (1994).
  [hep-ph/9310379].



\bibitem{Dashen:1994qi}
  R.~F.~Dashen, E.~E.~Jenkins, A.~V.~Manohar,
  Phys.\ Rev.\  {\bf D51}, 3697-3727 (1995).
  [hep-ph/9411234].



\bibitem{Jenkins:1995td}
  E.~E.~Jenkins, R.~F.~Lebed,
  Phys.\ Rev.\  {\bf D52}, 282-294 (1995).
  [hep-ph/9502227].

\bibitem{Dai:1995zg}
  J.~Dai, R.~F.~Dashen, E.~E.~Jenkins, A.~V.~Manohar,
  Phys.\ Rev.\  {\bf D53}, 273-282 (1996).
  [hep-ph/9506273].


\bibitem{Manohar:1998xv} 
  A.~V.~Manohar,
  ``Large N QCD,''
  hep-ph/9802419.


\bibitem{Wittig:2012ha} 
For a recent appraisal, see
  H.~Wittig,
  ``Low-energy particle physics and chiral extrapolations,''
  arXiv:1201.4774 [hep-lat].

\bibitem{Davies:2003ik}
  C.~T.~H.~Davies {\it et al.}  [HPQCD and UKQCD and MILC and Fermilab Lattice Collaborations],
  Phys.\ Rev.\ Lett.\  {\bf 92}, 022001 (2004)
  [hep-lat/0304004].


\bibitem{Sommer:1993ce}
  R.~Sommer,
  Nucl.\ Phys.\ B {\bf 411}, 839 (1994)
  [arXiv:hep-lat/9310022].


\bibitem{Bazavov:2009bb}
  A.~Bazavov, D.~Toussaint, C.~Bernard, J.~Laiho, C.~DeTar, L.~Levkova, M.~B.~Oktay and S.~Gottlieb {\it et al.},
  Rev.\ Mod.\ Phys.\  {\bf 82}, 1349 (2010)
  [arXiv:0903.3598 [hep-lat]].


\bibitem{MILC} {\tt http://www.physics.utah.edu/\%7Edetar/milc/}

\bibitem{DeGrand:2011qd} 
  T.~DeGrand, Y.~Shamir and B.~Svetitsky,
  Phys.\ Rev.\ D {\bf 83}, 074507 (2011)
  [arXiv:1102.2843 [hep-lat]];
  Phys.\ Rev.\ D {\bf 82}, 054503 (2010)
  [arXiv:1006.0707 [hep-lat]];
  arXiv:1202.2675 [hep-lat].


\bibitem{Brown:1987rra}
  F.~R.~Brown, T.~J.~Woch,
  Phys.\ Rev.\ Lett.\  {\bf 58}, 2394 (1987).
  
\bibitem{Cabibbo:1982zn}
  N.~Cabibbo, E.~Marinari,
  Phys.\ Lett.\  {\bf B119}, 387-390 (1982).





\bibitem{Hasenfratz:2007rf}
  A.~Hasenfratz, R.~Hoffmann, S.~Schaefer,
  JHEP {\bf 0705}, 029 (2007).
  [hep-lat/0702028].


\bibitem{DeGrand:2012qa} 
  T.~DeGrand, Y.~Shamir and B.~Svetitsky,
  arXiv:1202.2675 [hep-lat].

 

\bibitem{Kiskis:2003rd}
  J.~Kiskis, R.~Narayanan, H.~Neuberger,
  Phys.\ Lett.\  {\bf B574}, 65-74 (2003).
  [hep-lat/0308033].

\bibitem{deForcrand:2005xr}
  P.~de Forcrand, O.~Jahn,
    [hep-lat/0503041].

\bibitem{Creutz:1987xi}
  M.~Creutz,
  Phys.\ Rev.\  {\bf D36}, 515 (1987).
  

\end{thebibliography}
\end{document}